\documentclass[10pt,prl,twocolumn,showpacs,superscriptaddress,floatfix]{revtex4-2}
\usepackage{graphicx}
\usepackage{amssymb}
\usepackage{amsmath}
\usepackage{amsthm}
\usepackage{bm}
\usepackage{physics}
\usepackage{color,xcolor}
\usepackage{subfigure}
\usepackage{algpseudocode}
\usepackage{algorithm}
\usepackage{algorithmicx}
\usepackage{lineno}
\usepackage{soul}
\usepackage{mathtools} 
\usepackage{mathrsfs}
\usepackage{graphicx}% Include figure files
\usepackage{dcolumn}% Align table columns on decimal point
\usepackage{textcomp,mathcomp}
\usepackage{enumerate}
\newtheorem{theorem}{Theorem}

\usepackage{lipsum} % 用于生成虚拟文本，可以删除
\usepackage[colorlinks,
linkcolor=blue,      
anchorcolor=blue, 
citecolor=purple]{hyperref}

\begin{document}

\preprint{APS/123-QED}

\title{Simultaneous Heisenberg-Limited Multiparameter Metrology via Indefinite Evolution}% Force line breaks with \\

\author{Hang Xu}
\affiliation{State Key Laboratory of Photonics and Communications, Institute for Quantum Sensing and Information Processing, Shanghai Jiao Tong University, Shanghai 200240, People's Republic of China}%

\author{Tailong Xiao}%
\email{tailong\_shaw@sjtu.edu.cn}
\affiliation{State Key Laboratory of Photonics and Communications, Institute for Quantum Sensing and Information Processing, Shanghai Jiao Tong University, Shanghai 200240, People's Republic of China}%
 \affiliation{Hefei National Laboratory, Hefei, 230088, People's Republic of China}
\affiliation{Shanghai Research Center for Quantum Sciences, Shanghai, 201315, People's Republic of China}

\author{Ze Zheng}
\affiliation{State Key Laboratory of Photonics and Communications, Institute for Quantum Sensing and Information Processing, Shanghai Jiao Tong University, Shanghai 200240, People's Republic of China}%

\author{Xiaoyang Deng}
\affiliation{State Key Laboratory of Photonics and Communications, Institute for Quantum Sensing and Information Processing, Shanghai Jiao Tong University, Shanghai 200240, People's Republic of China}%

\author{Jinfeng Zheng}
\affiliation{State Key Laboratory of Photonics and Communications, Institute for Quantum Sensing and Information Processing, Shanghai Jiao Tong University, Shanghai 200240, People's Republic of China}%

\author{Jingzheng Huang}%
\affiliation{State Key Laboratory of Photonics and Communications, Institute for Quantum Sensing and Information Processing, Shanghai Jiao Tong University, Shanghai 200240, People's Republic of China}%
 \affiliation{Hefei National Laboratory, Hefei, 230088, People's Republic of China}
\affiliation{Shanghai Research Center for Quantum Sciences, Shanghai, 201315, People's Republic of China}

 \author{Guihua Zeng}%
 \email{ghzeng@sjtu.edu.cn}
\affiliation{State Key Laboratory of Photonics and Communications, Institute for Quantum Sensing and Information Processing, Shanghai Jiao Tong University, Shanghai 200240, People's Republic of China}%
 \affiliation{Hefei National Laboratory, Hefei, 230088, People's Republic of China}
\affiliation{Shanghai Research Center for Quantum Sciences, Shanghai, 201315, People's Republic of China}

\date{\today}

\begin{abstract}
%%%%%%%%%%
Quantum metrology achieves Heisenberg-limited precision in single-parameter estimation, but its multiparameter extension is fundamentally constrained by both parameter-encoding and measurement incompatibility. 
Noncommuting signal generators may cause incompatible parameter-encoding, preventing the quantum Fisher information matrix from simultaneously achieving the Heisenberg scale for all parameters. 
Due to incompatible optimal measurements, the classical Fisher information matrix represents the practical attainable precision. 
Here, we introduce a multiparameter metrology framework based on indefinite evolution (IE), in which different control operations and signal reversal are placed in a coherent superposition. 
For a single-qubit probe with mutually orthogonal signal generators, IE enables compatible parameter encoding and optimal measurement without the signal reversal.
For parallel generators, where only signal reversal realized by its generator is available, IE can achieve the same performance. 
% We further extend this mechanism to locally encoded many-body probes, high-dimensional probes, and noisy systems. 
We further extend this mechanism to noisy, many-body, and high-dimensional probes, and establish general conditions for achieving the simultaneous Heisenberg-limit.
In contrast, definite evolution cannot achieve the same performance under compatible optimal measurements, even when signal reversal is available. 
Our results identify IE as an operational resource for overcoming multiparameter incompatibility and open a route toward attainable Heisenberg-limited sensing in interferometric platforms.

\end{abstract}

\maketitle
\textit{Introduction.}---Quantum sensors \cite{QS1,QS2,QS3} can exploit coherence, entanglement \cite{echo1,echo2,echo3}, and controlled dynamics \cite{control1,control2} to achieve Heisenberg-limited precision in single-parameter estimation \cite{HL1,HL2,HL3,HL4,HL5,cai1,cai2,cai3,yuanxiao1,heqiong2}, thereby surpassing the standard quantum limit. Although environmental noise can severely degrade this quantum advantage \cite{noisysensing1,noisysensing4,nogo}, quantum error correction \cite{qec1,qec2,qec3,qec4,qec5,qec6,qec7,qec8,qec9,zss1}, error mitigation \cite{qem1,qem2}, and postselection \cite{postselect1} have shown that Heisenberg-limited scaling can still be restored or approximated in several restricted settings. Extending this advantage to multiparameter estimation, however, is far from straightforward. Multiparameter quantum metrology requires not only enhancing the information about each parameter, but also extracting such information simultaneously udner a compatible optimal measurement scheme \cite{multi1,multi2,multi3}.

The difficulty arises at two distinct levels. First, at the encoding level, the signal generators associated with different parameters generally do not commute. As a result, the quantum state cannot simultaneously follow the optimal encoding trajectory for each parameter, and the quantum Fisher information matrix (QFIM) itself may fail to exhibit Heisenberg-limit scaling for all parameters at once. Second, at the measurement level, the optimal measurements for different parameters are generally incompatible \cite{multi4,multi5}. Therefore, although the QFIM defines the quantum Cramér--Rao bound \cite{qfi1,qfi2}, it does not necessarily characterize the precision attainable in practice. For a given measurement, the precision is described by the classical Fisher information matrix (CFIM). After optimizing over measurements for a given weight matrix, the attainable precision is more appropriately characterized by the Holevo bound \cite{Holevo1}. These observations show that the simultaneous attainability of the Heisenberg limit in multiparameter metrology is constrained by both encoding incompatibility and measurement incompatibility \cite{multi6,multi7,multi8,multi9,multi10,multi11,multi12}.

Indefinite evolution (IE) offers a new route to overcome these limitations. Originating from the idea of quantum superpositions of different spacetime structures or causal orders \cite{gravity1,gravity2}, it allows a quantum system to coherently pass through distinct evolution channels. Resources such as indefinite causal order \cite{SWITCH1,SWITCH2,SWITCH3,SWITCH4,ICO1,ICO5} and indefinite time direction \cite{ITD1,ITD2,ITD3,ITD4,ITD5,ITD6} have recently been explored for enhancing quantum information processing \cite{ico_qcommu1,ico_qcommu2}, thermodynamic tasks \cite{ico_qt1,ico_qt2}, and quantum computing \cite{ico_qc1,ico_qc2}. In single-parameter metrology, indefinite evolution can coherently superpose different evolution paths, control operations, or Hamiltonian-sign configurations, which has been proven to improve measurement precision \cite{heqiong1,ico_cvsensing1,ico_cvsensing2} and suppress noise \cite{ico_dvsensing1,ico_dvsensing2}. However, its role in multiparameter estimation remains largely unexplored. 
% This naturally raises the central question: can indefinite evolution mitigate both encoding and measurement incompatibility, enabling multiple parameters to attain Heisenberg-limited precision simultaneously under a common measurement scheme?

% In this Letter, we develop a multiparameter metrology strategy based on indefinite evolution (IE), where a probe coherently traverses distinct channels controlled by an auxiliary system. These channels may contain different control operations and controlled sign reversals of the signal generators. By choosing auxiliary gates that anticommute with selected generators, path interference remaps the corresponding parameter information onto auxiliary degrees of freedom while preserving the remaining information in the probe. We show that, for qubit probes with orthogonal signal generators, this mechanism achieves the simultaneous Heisenberg limit under compatible optimal measurements without signal reversal, whereas for parallel generators controlled signal reversal is necessary. 
% We also prove that definite evolution cannot achieve the same simultaneous limit even with signal reversal.
% Finally, we extended the framework to noisy open systems and high-dimensional probes. In the former case, a cascade of multiple IE evolution schemes is required, while in the latter case, subspace projection becomes a useful operation.

In this paper, we develop a multiparameter metrology protocol based on IE, in which a probe can coherently undergo different channels controlled by auxiliary systems. These channels may involve different control operations and sign reversal of the signal generator.
By utilizing gate operations or signal reversal (SR), IE can map different signals from short-time evolutions to the auxiliary system and the probe, respectively. 
Through history-dependent compensatory control, the two signals can continuously accumulate in two subspaces, thereby achieving the simultaneous Heisenberg limit (SHL). 
% We demonstrate that, for a qubit probe with orthogonal signal generators, this mechanism achieves SHL under compatible optimal measurements without SR.
% In contrast, parallel generators require SR. 
% We also prove that even with signal reversal, definite evolution cannot achieve SHL. 
We prove that, for a qubit probe with orthogonal or non-parallel signal generators, this mechanism can achieve SHL under compatible optimal measurements without SR. 
However, for parallel signal generators, SR is further required.
By comparison, even with SR, definite evolution cannot achieve SHL.
Finally, we extend this framework to open systems and high-dimensional probes. For the former, this requires combining with IE-based error-correction protocols.
For the latter, we combine IE with subspace projection to confine the sensing dynamics to a two-dimensional encoding subspace. 
We further establish a general theorem showing that any pair of nonparallel traceless projected generators within this subspace can retain simultaneous HL scaling.

\textit{Multiparameter quantum sensing.}---Consider a multiparameter unitary sensing model in which the unknown parameters 
$\bm{h}=(h_1,h_2,\cdots,h_n)$ are encoded through $H=\sum_{i=1}^n h_i G_i$, where $G_i$ is the bare generator of $h_i$. The probe evolves as $|\psi_{\bm h}\rangle=U_{\bm h}|\psi_0\rangle$ with $U_{\bm h}=e^{-iH(\bm h)t}$.
For any locally unbiased estimator $\hat{\bm h}$ obtained from $M$ repetitions,
%%%%%%%%%%%%%%%%%%%%%%%%%%%%
\begin{equation}
{\rm Cov}(\hat{\bm h})\ge F^{-1}/M\ge Q^{-1}/M ,
\end{equation}
%%%%%%%%%%%%%%%%%%%%%%%%%%%%
where $F$ is the CFIM for a chosen measurement and $Q$ is the QFIM. For pure
states,
%%%%%%%%%%%%%%%%%%%%%%%%%%%%
\begin{equation}
\begin{split}
   &  Q_{ij}=4\,{\rm Cov}_{\psi_0}(\mathcal G_i,\mathcal G_j),\\
& \mathcal G_i=i(\partial_i U_{\bm h}^{\dagger})U_{\bm h}
=\int_0^t ds\, e^{iH(\bm h)s}G_i e^{-iH(\bm h)s}.
\end{split}
\end{equation}
Here $\mathcal G_i$ is the dressed generator associated with $h_i$. When
$[H,G_i]=0$, it reduces to $\mathcal G_i=tG_i$.
%%%%%%%%%%%%%%%%%%%%%%%%%%%%
This form highlights the key difficulty of multiparameter sensing. Unlike the
single-parameter case, noncommuting generators make the dressed generators
depend on the full Hamiltonian, and the measurements optimal for different
parameters are generally incompatible. Therefore, HL scaling of the QFIM alone
does not guarantee simultaneous attainability. Achieving the simultaneous
Heisenberg limit requires a common measurement whose CFIM also reaches HL
scaling for all parameters.

%%%%%%%%%%%%%%%%%%%%%%%%%%%%%%%%%%%%%%%%%%%%%%%
%%%%%%%%%%%%%%%%%%%%%%%%%%%%%%%%%%%%%%%%%%%%%%%%
\begin{figure}[htbp]
  \centering
\includegraphics[width=1\linewidth]{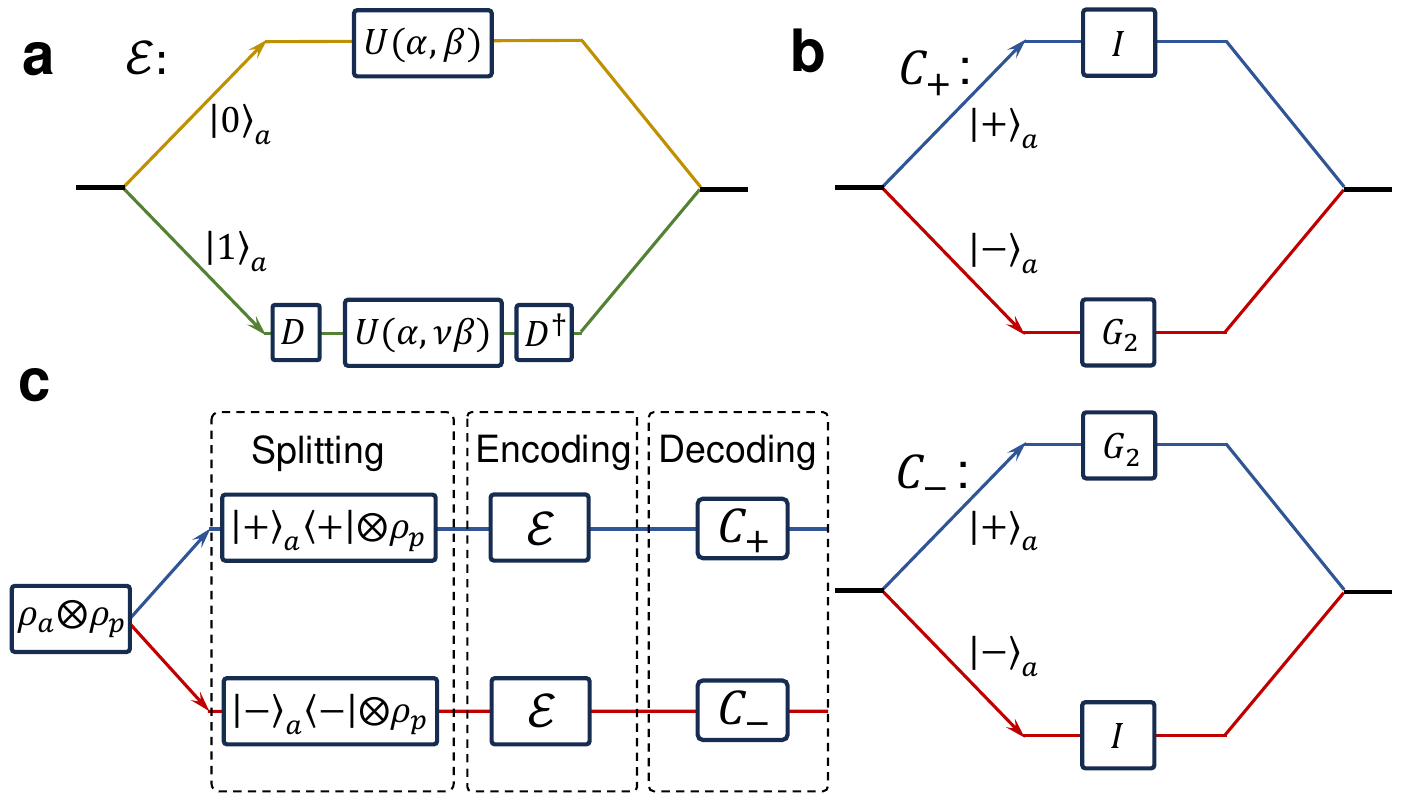}
  \caption{
Indefinite-evolution (IE) sensing protocol. 
(a) Encoding block $\mathcal{E}$.
% where the auxiliary qubit coherently controls the branches $U(\alpha,\beta)$ and $D U(\alpha,\nu\beta)D^\dagger$. 
(b) Conditional compensation blocks $C_{\pm}$ in the $|\pm\rangle_a$ basis. 
(c) Full IE time-slice protocol, composed of splitting, encoding, and history-dependent decoding.
}
\label{fig1}
\end{figure}

\textit{Indefinite-evolution sensing protocol.}---We consider a two-parameter unitary sensing dynamics
%%%%%%%%%%%%%%%%%%%%%%%%%%%%
\begin{equation}
U_t=e^{-i(\alpha G_1+\beta G_2)t},
\qquad
G_1^2=G_2^2=I .
\end{equation}
%%%%%%%%%%%%%%%%%%%%%%%%%%%%
The goal is to separate the two signal contributions into different degrees of freedom, so that they can be read out by compatible optimal measurements. 
The elementary encoding block is shown in Fig.~\ref{fig1}(a). 
The upper branch undergoes $U(\alpha,\beta)$, while the lower branch undergoes $D U(\alpha,\nu\beta)D^\dagger$, where $D$ is an auxiliary gate and $\nu=\pm1$ specifies whether the $\beta$ signal is reversed.

We first consider the case where SR is unavailable, i.e., $\nu=1$.
In the short-time limit ($\Delta t\ll1$), the sensing channel can be decomposed as $U_{\Delta t}(\alpha,\beta)
\simeq
e^{-i\beta G_2\Delta t}
e^{-i\alpha G_1\Delta t}$.
If there exists an auxiliary gate such that 
%%%%%%%%%%%%%%%%%%%%%%%%%%%%
\begin{equation}
[D,G_1]=0; \quad \{D,G_2\}=0,
\label{condition1}
\end{equation}
%%%%%%%%%%%%%%%%%%%%%%%%%%%%
for an auxiliary input $|+\rangle_a$ and a probe input $|\varphi\rangle_p$, the encoding block gives, % up to $O(\Delta t^2)$,
% If there exists an auxiliary gate such that $[D,G_1]=0$ and $\{D,G_2\}=0$, then the lower branch becomes
% %%%%%%%%%%%%%%%%%%%%%%%%%%%%
% \begin{equation}
% D U_{\Delta t}(\alpha,\beta)D^\dagger
% \simeq
% e^{+i\beta G_2\Delta t}
% e^{-i\alpha G_1\Delta t}.
% \end{equation}
% %%%%%%%%%%%%%%%%%%%%%%%%%%%%
% Therefore, the two branches differ only by the sign of the $\beta$ encoding, while the $\alpha$ encoding is identical in both branches. 
% For an auxiliary input $|+\rangle_a$ and a probe input $|\varphi\rangle_p$, the encoding block gives
%%%%%%%%%%%%%%%%%%%%%%%%%%%%
\begin{equation}
{{\cal E}_{\Delta t}} \circ |+\rangle_a|\varphi {\rangle _p} = {e^{ - i\beta {Z_a}{G_2}\Delta t}}{e^{ - i\alpha {G_1}\Delta t}}|+\rangle_a|\varphi {\rangle _p}.
\label{eq:ie_encoding_noSR}
\end{equation}
%%%%%%%%%%%%%%%%%%%%%%%%%%%%
For example, when $G_1=X$ and $G_2=Z$, choosing $D=X$ satisfies this condition.
Then the decoding block removes the residual $G_2$ on the probe space. 
Equivalently,
%%%%%%%%%%%%%%%%%%%%%%%%%%%%
\begin{equation}
C_+ \circ\mathcal{E}_{\Delta t} \circ
|+\rangle_a|\varphi\rangle_p
=
e^{-i\beta Z_a\Delta t}|+\rangle_a
\otimes
e^{-i\alpha G_1\Delta t}|\varphi\rangle_p .
\label{eq:ie_step}
\end{equation}
%%%%%%%%%%%%%%%%%%%%%%%%%%%%
Here $Z_a$ denotes the effective Pauli operator in the $|\pm\rangle_a$ basis.

A finite sensing time requires repeated time slices. 
After one slice, the auxiliary state already carries the accumulated $\beta$ phase, so the compensation in the next slice must depend on the incoming auxiliary branch. 
The $|+\rangle_a$ and $|-\rangle_a$ components are therefore decoded by two history-dependent compensation operations $C_+$ and $C_-$, respectively, as shown in Fig.~\ref{fig1}. 
This ensures that every branch continues to accumulate the same effective auxiliary evolution.

Repeating the protocol and taking the continuous limit gives
%%%%%%%%%%%%%%%%%%%%%%%%%%%%
\begin{equation}
|\psi(t)\rangle
\simeq
e^{-i\beta Z_a t}|+\rangle_a
\otimes
e^{-i\alpha G_1 t}|\varphi\rangle_p .
\label{eq:ie_finite}
\end{equation}
%%%%%%%%%%%%%%%%%%%%%%%%%%%%
Hence the $\beta$ signal is stored in the auxiliary qubit, while the $\alpha$ signal remains in the probe. 
The two parameters are therefore read out from independent subsystems. 
As a result, their optimal measurements are compatible, and the CFIM is diagonal with HL scaling for both parameters.

We next consider the case where SR is available, i.e., $\nu=-1$. 
% In this case, the lower branch directly contains the reversed $\beta$ signal. 
Taking $D=I$, the encoding block has the same effect as Eq.~(\ref{eq:ie_encoding_noSR}): the $\alpha$ encoding is unchanged in both branches, while the $\beta$ encoding has opposite signs in the two branches. 
Therefore, after applying the same decoding operation, one IE time slice again gives Eq.~(\ref{eq:ie_step}). 
With the same history-dependent compensation between consecutive slices, the full indefinite-evolution circuit also has the same effective separated dynamics as Eq.~(\ref{eq:ie_finite}). 
Hence, when SR is available, the two signals are again separated into independent subsystems, yielding a diagonal CFIM with HL scaling for both parameters.

%%%%%%%%%%%%%%%%%%%%%%%%%%%%%%%%%%%%%%%%%%%%%%%%
%%%%%%%%%%%%%%%%%%%%%%%%%%%%%%%%%%%%%%%%%%%%%%%%
\begin{figure}[htbp]
  \centering
  \includegraphics[width=1\linewidth]{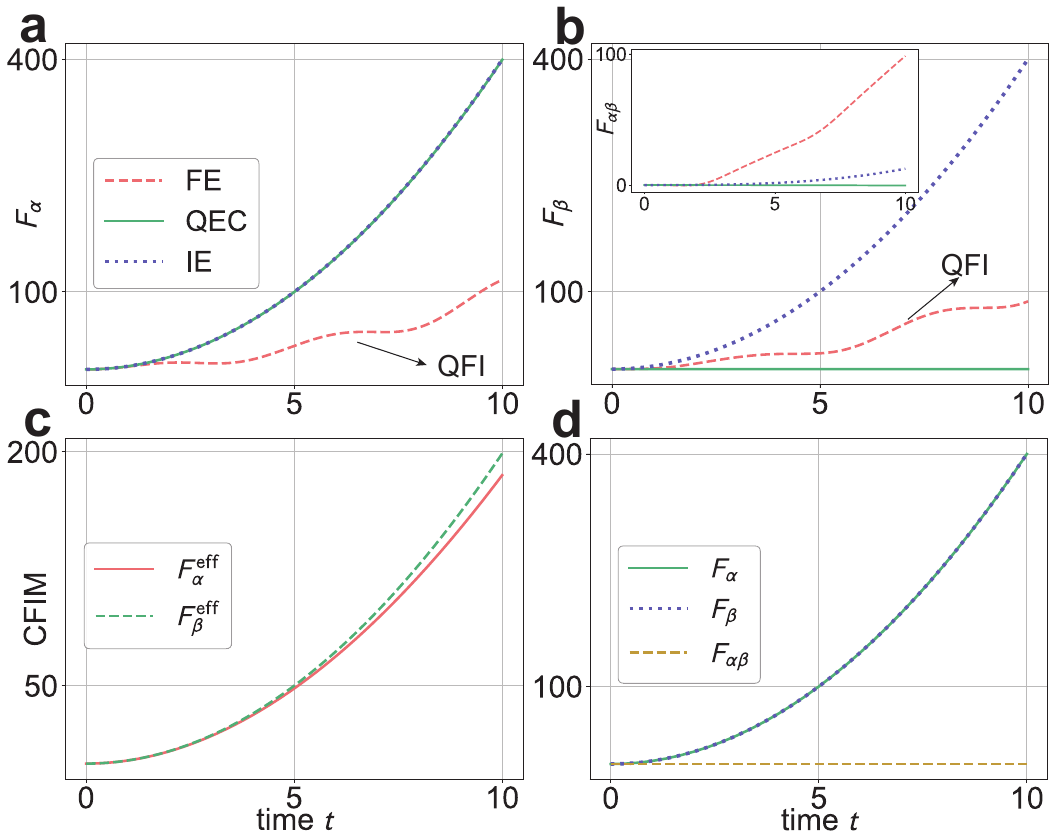}
  \caption{
  Results of the noiseless qubit probe. 
(a,b) Estimation precision for $\alpha$ and $\beta$ in the orthogonal case. 
(c) IE for the non-parallel case with $\theta=\pi/4$. 
(d) IE for the parallel case.
}
\label{fig2}
\end{figure}

\textit{Noiseless qubit probe.}---We first consider a noiseless qubit probe and compare three sensing strategies: free evolution (FE), a quantum-error-correction-inspired definite-evolution strategy (QEC), and IE. 
% In Fig.~\ref{fig2}, FE is characterized by the QFIM, while QEC and IE are characterized by the CFIM associated with explicit measurements.

We begin with the case of orthogonal signal generators $H=\alpha X+\beta Z$.
For FE, the evolution operator of the probe can be written explicitly as
% %%%%%%%%%%%%%%%%%%%%%%%%%%%%
% \begin{equation}
% U(t)
% =
% \cos(Bt)I
% -i\frac{\sin(Bt)}{B}(\alpha X+\beta Z),
% \end{equation}
% %%%%%%%%%%%%%%%%%%%%%%%%%%%%
%%%%%%%%%%%%%%%%%%%%%%%%%%%%
\begin{equation}
U(t)=\cos(Bt)I-i{\sin(Bt)}(\alpha X+\beta Z)/B,
\end{equation}
%%%%%%%%%%%%%%%%%%%%%%%%%%%%
where $B=\sqrt{\alpha^2+\beta^2}$.
% This form shows that the two parameters are encoded into the same rotation frequency $B$ and the same rotation axis $(\alpha,0,\beta)/B$. 
% As a result, changing one parameter also changes the effective encoding direction of the other. 
% The probe cannot simultaneously follow the optimal encoding path for both $X$ and $Z$ signals. 
% This is fundamentally different from two independent single-parameter evolutions, where the optimal generators would be $X$ and $Z$ separately. 
Consequently, FE cannot provide SHL for both parameters, as shown by the red dashed curves in Fig.~\ref{fig2}.
The QEC protocol represents a definite-evolution strategy that protects a selected signal component by continuously extracting a syndrome. 
In the present model, the QEC construction preserves the $\alpha$ signal and suppresses the $Z$ component as an error. 
Therefore, it can recover the HL for $F_{\alpha}$, as shown in Fig.~\ref{fig2}(a), but the syndrome extraction erases the information carried by $\beta Z$, leading to a vanishing or strongly suppressed $F_{\beta}$ in Fig.~\ref{fig2}(b). 
% This illustrates a central limitation of QEC-like definite evolution: protecting one parameter can remove the information of another.
In contrast, IE does not eliminate the $\beta$ signal. 
It uses coherent path interference to remap the $\beta$ information onto an auxiliary qubit while keeping the $\alpha$ information in the probe. 
% In the continuous-time limit, the output state takes the separated form
% %%%%%%%%%%%%%%%%%%%%%%%%%%%%
% \begin{equation}
% |\psi(t)\rangle
% \simeq
% e^{-i\beta Z_a t}|+\rangle_a
% \otimes
% e^{-i\alpha X t}|\varphi\rangle_p .
% \label{eq:ie_output_orthogonal}
% \end{equation}
% %%%%%%%%%%%%%%%%%%%%%%%%%%%%
Figure~\ref{fig2}(a) and (b) show that IE reaches the HL for both parameters. 
Thus, for orthogonal signal generators, IE achieves the SHL without SR.

We then consider the non-parallel signal generators, $H=\alpha X+\beta(X\sin\theta+Z\cos\theta)$.
For $0<\theta<\pi/2$, no auxiliary gate can satisfy Eq.~(\ref{condition1}) for the case where SR is unavailable. 
Nevertheless, even without SR, IE can still extract the $Z$ component of the $\beta$ signal into the auxiliary qubit, while the probe retains the parallel $X$ component. 
Combining the probe and auxiliary measurements gives
%%%%%%%%%%%%%%%%%%%%%%%%%%%%
\begin{equation}
F\simeq4t^2
\begin{pmatrix}
1 & \sin\theta \\
\sin\theta & 1
\end{pmatrix}.
\label{eq:nonorthogonal_total_cfim}
\end{equation}
%%%%%%%%%%%%%%%%%%%%%%%%%%%%
Although $F$ is not diagonal, the effective CFI~\cite{EffCFI} obtained from the inverse CFIM gives $F_{\alpha}^{\rm eff}\simeq F_{\beta}^{\rm eff}\simeq4t^2\cos^2\theta$, as shown in Fig.~\ref{fig2}(c).
Thus the IE protocol still achieves simultaneous HL scaling for non-parallel signal generators without SR.

We finally consider the parallel signal generators, $H=(\alpha+\beta)Z$.
In FE, only the sum $\alpha+\beta$ is encoded in the probe, so the two parameters cannot be identified independently. 
Equivalently, the CFIM is singular and the variance of at least one independent parameter combination diverges. 
Without SR, IE cannot distinguish the two parameters either. 
To recover identifiability and achieve the SHL, one must introduce SR in the encoding phase. 
% By arranging different spatial branches to experience opposite signs of one signal generator while leaving the other unchanged, IE makes $\alpha$ and $\beta$ physically distinguishable and maps them to different degrees of freedom. 
As shown in Fig.~\ref{fig2}(d), the resulting CFIM becomes nonsingular: both $F_{\alpha}$ and $F_{\beta}$ exhibit Heisenberg scaling, while the cross term $F_{\alpha\beta}$ remains close to zero. 
Therefore, for parallel signal generators, SR is a necessary resource for IE to achieve the SHL.

%%%%%%%%%%%%%%%%%%%%%%%%%%%%
\begin{table*}[t]
\caption{
Precision limits for noiseless qubit sensing. 
% DE denotes definite evolution and IE denotes indefinite evolution. 
% HL denotes Heisenberg-limit scaling for a single parameter, while SHL denotes simultaneous Heisenberg-limit scaling for both parameters under a common measurement.
}
\label{tab:qubit_summary}
\begin{ruledtabular}
\begin{tabular}{lccc}
Signal relation & SR & DE & IE \\
\hline
$G_1\neq G_2$
& unavailable
& no SHL; at most single-parameter HL
& SHL \\
$G_1\neq G_2$
& available
& single-parameter HL by temporal reversal
& SHL \\
$G_1 = G_2$
& unavailable
& singular CFIM; non-identifiable
& singular CFIM; non-identifiable \\
$G_1 = G_2$
& available
& single-parameter HL by temporal reversal
& SHL \\
\end{tabular}
\end{ruledtabular}
\end{table*}
%%%%%%%%%%%%%%%%%%%%%%%%%%%%

Finally, we clarify the role of SR in definite evolution (DE). 
In DE, SR can only be applied sequentially in time. 
Consider the two short-time channels
%%%%%%%%%%%%%%%%%%%%%%%%%%%%
\begin{equation}
U_{\Delta t/2}
=
e^{-i(\alpha G_1+\beta G_2)\Delta t/2},
U'_{\Delta t/2}
=
e^{-i(\alpha G_1-\beta G_2)\Delta t/2}.
\label{eq:temporal_reversal_channels}
\end{equation}
%%%%%%%%%%%%%%%%%%%%%%%%%%%%
Their sequential application gives
%%%%%%%%%%%%%%%%%%%%%%%%%%%%
\begin{equation}
{({U'_{\Delta t/2}}{U_{\Delta t/2}})^N} \simeq {e^{ - i\alpha G_1t}},\;\;N = t/\Delta t.
\label{eq:temporal_reversal_total}
\end{equation}
%%%%%%%%%%%%%%%%%%%%%%%%%%%%
Thus, temporal SR cancels the $\beta$ signal instead of transferring it to an independent degree of freedom. 
DE can at most recover the HL for one parameter at the cost of removing the other. 
In the parallel case, temporal SR can remove the CFIM singularity, but it still yields a single-parameter HL. 
The precision limits of single qubit probe are summarized in Table~\ref{tab:qubit_summary}.

%%%%%%%%%%%%%%%%%%%%%%%%%%%%%%%%%%%%%%%%%%%%%%%%
%%%%%%%%%%%%%%%%%%%%%%%%%%%%%%%%%%%%%%%%%%%%%%%%
\begin{figure}[htbp]
  \centering
  \includegraphics[width=1\linewidth]{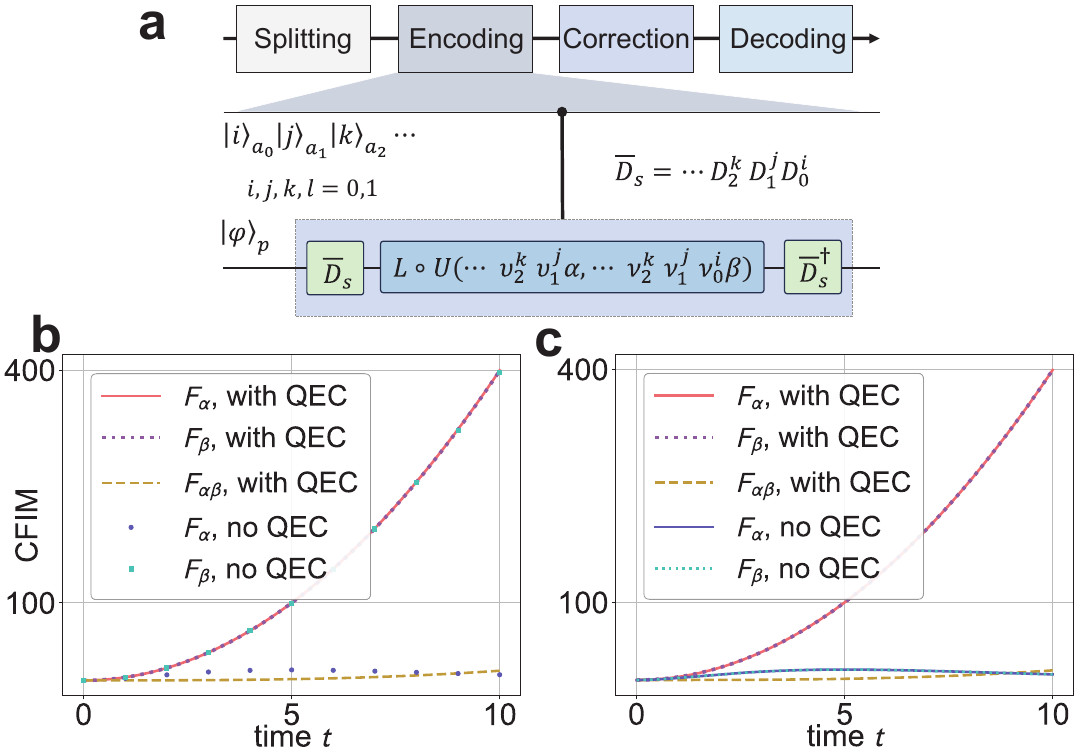}
  \caption{
  (a) Schematic of the noisy IE protocol. The evolution channel  of the probe in encoding phase is controlled by the computational basis of the auxiliary qubits.
  Results of the noisy qubit probe by IE protocol for single $X$-noise (b), $X$- and $Z$-noises case (c). 
}
  \label{fig3}
\end{figure}

\textit{Noisy qubit probe.}---We consider the noisy qubit probe and focus on the IE protocol. 
The signal Hamiltonian is still $H=\alpha X+\beta Z$, while the probe is subject to Markovian noise during the sensing process.

We first consider the case with a single $X$-type noise. 
Because the noise operator is parallel to the $\alpha$ signal, additional SR-assisted syndrome extraction is required to separate the signal and noise contributions (see the Supplementary Materials). 
As shown in Fig.~\ref{fig3}(a), compared to the noise-free IE protocol, the noisy IE protocol introduces an additional auxiliary qubit during the encoding phase to detect the noise, followed by syndrome-dependent recovery. 
Figure~\ref{fig3}(b) shows that, with noise correction, the CFIM elements $F_{\alpha}$ and $F_{\beta}$ both achieve HL, and the cross term $F_{\alpha\beta}$ remains close to zero. 
Without noise correction, the $X$ noise mainly destroys the precision of the $\alpha$ signal, while the $\beta$ signal stored in the auxiliary space is unaffected.

We further consider the case with both $X$- and $Z$-type noises. 
In this case, each noise component is parallel to one signal component, and three auxiliary qubits are used: one for receiving the $\beta$ signal and two for extracting the $X$- and $Z$-noise syndromes. 
Figure~\ref{fig3}(c) shows that, after noise correction, the IE protocol restores HL for both $F_{\alpha}$ and $F_{\beta}$ with negligible $F_{\alpha\beta}$. 
In contrast, without noise correction, both signals are degraded by the corresponding noise channels. 
These results show that IE can be combined with noise-syndrome extraction to preserve the SHL in noisy multiparameter sensing, provided that SR resources are available.

\textit{Multi-qubit probe.}---We now extend the IE protocol to a noiseless multi-qubit probe. 
Here we focus on the orthogonal-signal case. 
The signal Hamiltonian is
%%%%%%%%%%%%%%%%%%%%%%%%%%%%
\begin{equation}
H=\alpha S_x^{p}+\beta S_z^{p},
\;
S_x^{p}=\sum_{j=1}^{n}X_j^{p},
\;
S_z^{p}=\sum_{j=1}^{n}Z_j^{p}.
\label{eq:many_qubit_hamiltonian}
\end{equation}
%%%%%%%%%%%%%%%%%%%%%%%%%%%%
where $n$ is the number of probe qubits. 
Since the signal is locally encoded on each probe qubit, the single-qubit IE protocol can be applied in parallel. 
Specifically, each probe qubit is equipped with one auxiliary qubit, which coherently controls the local evolution branch of that probe. 
The local IE operation separates the two noncommuting signal components by mapping the $Z_j$ signal to the corresponding auxiliary qubit while retaining the $X_j$ signal in the probe.

\begin{figure}[htbp]
  \centering
  \includegraphics[width=1\linewidth]{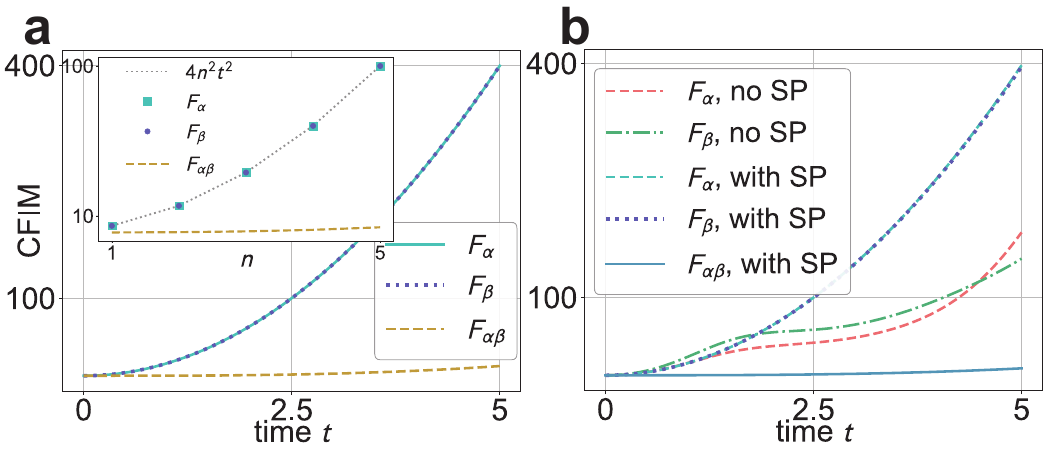}
  \caption{
(a) Multi-qubit probe for $n=2$; the inset shows the qubit-number $n$ scaling at $t=1$. 
(b) High-dimensional probe with and without SP.
}
  \label{fig4}
\end{figure}

In the continuous-time limit, the multi-qubit IE protocol gives the effective separated evolution
%%%%%%%%%%%%%%%%%%%%%%%%%%%%
\begin{equation}
|\Psi(t)\rangle
\simeq
e^{-i\beta S_z^{a}t}|\Psi_a(0)\rangle
\otimes
e^{-i\alpha S_x^{p}t}|\Psi_p(0)\rangle.
\label{eq:many_qubit_ie_output}
\end{equation}
%%%%%%%%%%%%%%%%%%%%%%%%%%%%
% Thus the $\beta$ signal is collectively stored in the auxiliary register, while the $\alpha$ signal remains collectively encoded in the probe register. 
The numerical results are shown in Fig.~\ref{fig4}(a). 
For $n=2$, both $F_{\alpha}$ and $F_{\beta}$ follow the ideal scaling $4(nt)^2$ over the sensing time, while the cross term $F_{\alpha\beta}$ remains zero. 
The inset further shows the qubit-number scaling at a fixed sensing time $t=1$, where both $F_{\alpha}$ and $F_{\beta}$ scale as $n^2$, confirming that the IE protocol achieves the HL on the spatial scale for both parameters simultaneously.

\textit{High-dimensional probe.}---We consider the extension of IE to a high-dimensional probe.
We diagonalize $G_1$ and choose two nondegenerate eigenstates
$|\lambda_1\rangle$ and $|\lambda_2\rangle$ with the largest eigenvalue separation, defining
%%%%%%%%%%%%%%%%%%%%%%%%%%%%
\begin{equation}
\mathcal{H}_{\rm enc}={\rm span}\{|\lambda_1\rangle,|\lambda_2\rangle\},
\;
P=|\lambda_1\rangle\langle\lambda_1| +|\lambda_2\rangle\langle\lambda_2| .
\end{equation}
%%%%%%%%%%%%%%%%%%%%%%%%%%%%
If $G_2$ is block diagonal with respect to
$\mathcal{H}=\mathcal{H}_{\rm enc}\oplus\mathcal{H}_{\rm enc}^{\perp}$,
the dynamics is directly reduced to a two-level model.
Otherwise, the leakage induced by $G_2$ can be suppressed by frequent subspace projections (SPs), yielding the effective generators $P_i=PG_i P$ in the Zeno limit~\cite{SPZeno}.
The sensing-relevant projected generators can be represented as
$\widetilde G_i=\mathbf g_i\cdot\boldsymbol{\sigma}$ after removing their identity components~\cite{ProjectedGenerator}.

\begin{theorem}\label{the1}
% \textbf{Theorem 1.}---
If $\mathbf g_1$ and $\mathbf g_2$ are nonparallel, IE without SR simultaneously retains HL scaling for both parameters.
The corresponding effective CFIs are
%%%%%%%%%%%%%%%%%%%%%%%%%%%%
\begin{equation}
F_{\alpha}^{\rm eff}
=
4t^2|\mathbf g_1|^2\sin^2\theta,
\qquad
F_{\beta}^{\rm eff}
=
4t^2|\mathbf g_2|^2\sin^2\theta,
\label{eq:highdim_effective_cfi}
\end{equation}
%%%%%%%%%%%%%%%%%%%%%%%%%%%%
where $\theta$ is the angle between $\mathbf g_1$ and $\mathbf g_2$~\cite{HighDimTheorem}.
Thus, nonparallel projected generators are sufficient for simultaneous HL scaling; orthogonality ($\theta=\pi/2$) recovers the full single-parameter HLs, whereas the effective CFIs vanish in the parallel limit ($\theta=0$).
\end{theorem}

As an example, we consider the three-level probe
%%%%%%%%%%%%%%%%%%%%%%%%%%%%
\begin{equation}
\begin{aligned}
G_1
&=
-2|0\rangle\langle0|
+
2|1\rangle\langle1|
+
|2\rangle\langle2|,
\\
G_2
&=
2\left(
|0\rangle\langle1|
+
|1\rangle\langle0|
+
|1\rangle\langle2|
+
|2\rangle\langle1|
\right).
\end{aligned}
\end{equation}
%%%%%%%%%%%%%%%%%%%%%%%%%%%%
The optimal encoding subspace is
$\mathcal{H}_{\rm enc}={\rm span}\{|0\rangle,|1\rangle\}$.
The projected generators are orthogonal with
$|\mathbf g_1|=|\mathbf g_2|=2$, giving
$F_{\alpha}^{\rm eff}=F_{\beta}^{\rm eff} \simeq 16t^2$~\cite{ThreeLevelProjection}.
As shown in Fig.~\ref{fig4}(b), SP therefore restores the SHL, whereas without SP, leakage from $\mathcal{H}_{\rm enc}$ degrades the CFIM.
%%%%%%%%%%%%%%%%%%%%%%%%%%%%%%%%%%%%%%%%%%%%%%%%
%%%%%%%%%%%%%%%%%%%%%%%%%%%%%%%%%%%%%%%%%%%%%%%%
%%%%%%%%%%%%%%%%%%%%%%%%%%%%%%%%%%%%%%%%%%%%%%%%
%%%%%%%%%%%%%%%%%%%%%%%%%%%%%%%%%%%%%%%%%%%%%%%%

\textit{Conclusions.}---In summary, we have introduced an indefinite evolution (IE) protocol for multiparameter quantum metrology. 
By coherently controlling different sensing evolutions through auxiliary systems, the IE directs different signal components to different degrees of freedom, thereby extracting them via compatible optimal measurements. 
This mechanism overcomes the incompatibility of encoding and measurement, making the simultaneous Heisenberg limit (SHL) achievable. 
For noise-free qubit probes, orthogonal signal generators can separate signals without signal reversal. 
In contrast, definite evolution can only use signal reversal in a time-sequential manner, which simply erases the accumulated signal. 
We further demonstrate that IE can detect and correct noise using additional auxiliary qubits. 
Moreover, the protocol can be extended to locally encoded multi-qubit probes to achieve the Heisenberg limit on the spatial scale, as well as to high-dimensional probes based on subspace projection. 
These results indicate that IE is a powerful resource for multiparameter quantum sensing.
More broadly, coherent information routing through indefinite evolution may provide a useful principle for quantum information processing beyond metrology. 
For quantum computing \cite{yuanxiao2,qqs1}, it suggests new ways to separate useful dynamics from control errors or noise, with potential relevance to programmable control and error-transparent gate design. 
For quantum communication and quantum networks  \cite{qn1}, it may offer a coherent strategy for distinguishing channel contributions, improving channel estimation, and routing information through auxiliary degrees of freedom.

\textit{Acknowledgments.}---The authors acknowledge support from the National Key R\&D Program of China (No.~2025YFF0515504), the National Natural Science Foundation of China (No.~62401359, 62471289), the State Key Laboratory of Photonics and Communications, the Quantum Science and Technology—National Science and Technology Major Project (No.~2021ZD0300703), and the Shanghai Municipal Science and Technology Major Project (No.~2019SHZDZX01).

\nocite{*}

\bibliography{sample}

\end{document}